\documentclass[oneside,a4paper,reqno]{amsart}
\usepackage{amssymb}
\usepackage{lscape}

\theoremstyle{remark}

\newcommand{\be}{\begin{equation}}
\newcommand{\ee}{\end{equation}}
\newcommand{\bc}{\begin{center}}
\newcommand{\ec}{\end{center}}
\newcommand{\bmt}{\begin{pmatrix}}
\newcommand{\emt}{\end{pmatrix}}

\DeclareMathOperator{\tr}{tr}
\DeclareMathOperator{\SL}{SL}

\DeclareMathOperator{\su}{su}

\begin{document}
\begin{titlepage}
\begin{center}
\bfseries  SIC-POVMS AND MUBS:  GEOMETRICAL RELATIONSHIPS IN PRIME DIMENSION
\end{center}
\vspace{1 cm}
\begin{center} D M APPLEBY
\end{center}
\begin{center} Department of Physics, Queen Mary
University of London,  Mile End Rd, London E1 4NS,
UK
 \end{center}
\vspace{0.5 cm}
\begin{center}
  (E-mail:  D.M.Appleby@qmul.ac.uk)
\end{center}
\vspace{0.75 cm}
\vspace{1.25 cm}
\begin{center}
\vspace{0.35 cm}
\parbox{12 cm }{ 
The paper concerns Weyl-Heisenberg covariant SIC-POVMs (symmetric informationally complete positive operator valued measures) and full sets of MUBs (mutually unbiased bases) in prime dimension.  When represented as vectors in generalized Bloch space a SIC-POVM forms a $d^2-1$ dimensional regular simplex ($d$ being the Hilbert space dimension).  By contrast, the generalized Bloch vectors representing a full set of MUBs form  $d+1$ mutually orthogonal $d-1$ dimensional regular simplices.  In this paper we show that, in the Weyl-Heisenberg case, there are some simple geometrical relationships between the single SIC-POVM simplex and the $d+1$ MUB simplices.  We go on to give geometrical interpretations of the minimum uncertainty states introduced by Wootters and Sussman, and by Appleby, Dang and Fuchs, and of the  fiduciality condition given by Appleby, Dang and Fuchs.
   }
\end{center}
\end{titlepage}
\section{Introduction}
There has been much interest in recent years in SIC-POVMs~\cite{Hoggar, Zauner, Caves, FuchsA,Renes, SanigaPlanatRosu,FuchsB,Rehacek,WoottersA,Ingemar1,GrasslA,Ingemar2,Koenig, ZimanBuzek,me1,KlappA,KlappB,GrasslB, Ballester,Gross,Colin,GodsilRoy,Scott,DurtA,Flammia,Kim,me2,RoyScott,KiblerA,ADF,Khat,Bodmann,KiblerB,Ingemar3} (symmetric informationally complete positive operator valued measures; citations in order of first appearance online or in print).  SIC-POVMs have been constructed analytically in Hilbert space dimension $d=2$--$13$, $15$ and $19$ (existence of analytic solutions for $d=11,15$ communicated to author privately~\cite{GrasslUnpub}; for $d=15$ also see ref.~\cite{meUnpub}), and numerically in dimension $5$--$45$.  The fact that they exist in every dimension up to $45$ (at least to a very high degree of numerical accuracy) means one may plausibly speculate that they exist in every dimension---although this remains to be proved.

MUBs~\cite{SanigaPlanatRosu,WoottersA, Ingemar1, GrasslA, Ingemar2,Ivanovic,WoottersB,WoottersC,Bandy,Pitt,KlappC,Archer,DurtB,WG,Planat, MubsD6}  (mutually unbiased bases; only a representative selection of  papers cited) have been the subject of intense investigation for a rather longer period of time.  It is known that the number of MUBs in dimension $d$ cannot exceed $d+1$, and that the maximum number of $d+1$ MUBs exist whenever $d$ is a power of a prime number.  However it remains an open question whether the maximal number of $d+1$ MUBs exist in any non-prime-power dimension.

The purpose of this paper is to describe some geometrical relationships between these somewhat elusive structures when the dimension $d$ is a prime number, so that full sets of $d+1$ MUBs certainly exist, and SIC-POVMs exist at least in some cases.    We also give a geometrical interpretation of the minimum uncertainty states introduced by Wootters and Sussman~\cite{WS,Suss} and Appleby, Dang and Fuchs~\cite{ADF}, and of the related fiduciality conditions given by Appleby, Dang and Fuchs~~\cite{ADF}.  Our analysis builds on previous work by Bengtsson and Ericcson~\cite{Ingemar1,Ingemar2} and Appleby~\cite{me2}. 

We will work in generalized Bloch space~\cite{Ingemar1,Ingemar2,me2,Harriman,Mahler,Jakobczyk,Kim1,Byrd,Schirmer,Kim2,Dietz}.  If $d=2$ the Bloch representation of an arbitrary density matrix  $\rho$ is defined by 
\be
\rho = \frac{1}{2} \left( 1 + \mathbf{b}\cdot\boldsymbol{\sigma}\right)
\ee 
where $\mathbf{b}$ is a $3$-vector having length $\le 1$ (the Bloch vector), and $\sigma_1$, $\sigma_2$, $\sigma_3$ are the Pauli matrices.    Pure states correspond to the case $|\mathbf{b}| = 1$, and mixed states to the case $|\mathbf{b}| < 1$. The crucial feature here is that $\mathbf{b}\cdot\boldsymbol{\sigma}$ is a trace-zero matrix.  So if $d>2$ we can proceed in an analogous manner, by writing 
\be
\rho = \frac{1}{d} \left( 1 + \mathbf{B}\right)
\ee
where $\mathbf{B}$ is a trace-zero Hermitian matrix acting on $d$-dimensional Hilbert Space $\mathcal{H}_d$.  More technically, $\mathbf{B}$ is an element of the Lie algebra $\su(d)$  (although we will make no use of the theory of Lie algebras in this paper).  To obtain the analogue of the Bloch vector $\mathbf{b}$ some authors proceed by choosing a basis for $\su(d)$ (for example, the Gell-Mann matrices).  We, however, find it more convenient to regard the matrix $\mathbf{B}$ itself as a vector, and to define an inner-product on $\su(d)$ by the formula
\be
\langle \mathbf{B} , \mathbf{B}' \rangle = \frac{1}{d(d-1)} \tr \left( \mathbf{B} \mathbf{B}' \right)
\ee
(readers familiar with the theory of Lie algebras will perceive that, apart from a scale factor, this is just the Killing form on $\su(d)$).  Note that this gives us a \emph{real} inner product space.  We define
\be
\| \mathbf{B} \| =  \sqrt{ \langle \mathbf{B}, \mathbf{B} \rangle }
\ee  
to be the corresponding norm.  

If $d=2$ the set of Bloch vectors corresponding to density matrices is very easily described:  it is just the ball $\| \mathbf{B} \| \le 1$.  If $d>2$ the geometry is rather more intricate. Let $\mathcal{B}$ be the Bloch body (the set of Bloch vectors corresponding to density matrices).  Then~\cite{Harriman,Kim1,Kim2} $\mathcal{B}$ is wholly contained by the out-ball 
\be
\mathcal{B}_{\mathrm{o}} = \{\mathbf{B} \in \su(d) \colon \| \mathbf{B} \| \le 1 \}
\ee
and wholly contains the in-ball
\be
\mathcal{B}_{\mathrm{i}} = \left\{\mathbf{B} \in \su(d) \colon \| \mathbf{B} \| \le 1/(d-1)\right\}
\ee
More succinctly:
\be
\mathcal{B}_{\mathrm{i}} \subseteq \mathcal{B} \subseteq \mathcal{B}_{\mathrm{o}}
\ee
Moreover $\mathcal{B}_{\mathrm{o}}$ and $\mathcal{B}_{\mathrm{i}}$ are, respectively, the smallest and largest balls with these properties. In other words
\begin{enumerate}
\item For each $0\le r \le 1$ there exists a Bloch vector $\mathbf{B}\in \mathcal{B}$ such that $\|\mathbf{B}\|=r$
\item For every $r > 1/(d-1)$ there exists a vector $\mathbf{B}$ such that $\|\mathbf{B}\| = r$ and $\mathbf{B} \notin \mathcal{B}$
\end{enumerate}
The set of pure states is the intersection of $\mathcal{B}$ with the out-sphere $\mathcal{S}_{\mathrm{0}}=\{\mathbf{B} \in \su(d) \colon \|\mathbf{B}\| = 1\}$.

In terms of this picture an orthonormal basis in the Hilbert space $\mathcal{H}_d$ corresponds to a $d-1$ dimensional regular simplex in Bloch space, wholly contained in the Bloch body and with all of its vertices lying on the out-sphere $\mathcal{S}_{\mathrm{o}}$.  To be more specific, let $|\psi_1\rangle, \dots , |\psi_d\rangle$ be an orthonormal basis in $\mathcal{H}_d$ and let $\mathbf{B}_a = d |\psi_a\rangle\langle\psi_a| -1$ be the corresponding Bloch vectors. Then
\begin{align}
\sum_{a=1}^{d} \mathbf{B}_a & = 0
\\
\intertext{and}
\langle \mathbf{B}_a , \mathbf{B}_b\rangle & = \frac{1}{d-1} \left(d \delta_{ab}-1\right)
\end{align}
Now let $|\psi'_1\rangle , \dots, |\psi'_d\rangle$ be a second orthonormal basis, and let $\mathbf{B}'_1, \dots , \mathbf{B}'_d$ be the corresponding Bloch vectors.  The two bases are mutally unbiased with respect to one another if and only if $\left|\langle \psi_a | \psi'_b \rangle\right| = 1/\sqrt{d}$ for all $a, b$.  In terms of Bloch space this means the bases are mutually unbiased if and only if the corresponding simplices are mutually orthogonal:
\be
\langle \mathbf{B}_a , \mathbf{B}'_b\rangle  = 0
\ee 
for all $a, b$.  It follows that a full set of $d+1$ MUBs~\cite{Ingemar1,Ingemar2,me2} corresponds to a family of $d+1$ mutually orthogonal $d-1$ dimensional simplices in Bloch space.  They thus define a highly symmetric polytope in Bloch space.  This picture also makes it clear~\cite{Ingemar1,Ingemar2,me2} why the question, whether full sets of $d+1$ MUBs exist in every dimension $d$, is hard to answer.  The problem is that the out-sphere $\mathcal{S}_{\mathrm{o}}$ has  dimension $d^2-2$, whereas the manifold of pure states $\mathcal{B}\cap\mathcal{S}_{\mathrm{o}}$ has dimension $2d-2$, which is smaller  if $d>2$, and very much smaller if $d$ is large.  It is easy to construct a family of $d+1$ mutually orthogonal $d-1$ dimensional simplices whose vertices lie on the out-sphere $\mathcal{S}_{\mathrm{o}}$, but if $d>2$ it is much less easy then to rotate this structure so that its vertices all lie on the lower dimensional sub-manifold $\mathcal{B}\cap\mathcal{S}_{\mathrm{o}}$.

A SIC-POVM~\cite{Ingemar1,Ingemar2,me2}, by contrast, corresponds to a single $d^2-1$ dimensional regular simplex in Bloch space, wholly contained in the Bloch body, and with its vertices all on the out-sphere $\mathcal{S}_{\mathrm{o}}$.  To be more specific, let $|\psi_1\rangle, \dots |\psi_{d^2}\rangle$ be the vectors defining a SIC-POVM.  So
\begin{align}
\frac{1}{d} \sum_{r=0}^{d^2} |\psi_r\rangle\langle \psi_r | & = 0
\\
\intertext{and}
\left|\langle \psi_r | \psi_s\rangle\right|^2 & = \frac{1}{d+1} \left(d \delta_{rs} + 1\right)
\end{align}
for all $r$, $s$.  In terms of the corresponding Bloch vectors $\mathbf{B}_r$ this means
\begin{align}
\sum_{r=0}^{d^2} \mathbf{B}_r & = 0
\\
\intertext{and}
\langle \mathbf{B}_r , \mathbf{B}_s \rangle & = \frac{1}{d^2-1} \left(d^2 \delta_{rs}-1\right)
\end{align}
for all $r, s$.  It is hard to prove that this structure exists in dimensions $>2$ for essentially the same reason that it is hard to prove  full sets of MUBs exist in dimensions $>2$:  namely~\cite{Ingemar1,Ingemar2,me2}, though it is easy to construct a $d^2-1$ dimensional regular simplex with its vertices all on the out-sphere $\mathcal{S}_{\mathrm{o}}$, it is very difficult then to rotate the simplex so that its vertices all lie on the lower dimensional sub-manifold $\mathcal{B}\cap\mathcal{S}_{\mathrm{o}}$.

The purpose of this paper is to examine the relationship between the $d^2-1$ dimensional simplex corresponding to a SIC-POVM, and the family of $d+1$ mutually orthogonal $d-1$ dimensional simplices corresponding to a full set of MUBs.  We will confine our attention to the case where $d$ is a prime number $>2$ (the case $d=2$ being almost trivial).  We will further restrict our attention to SIC-POVMs and MUBs which are covariant under the action of the Weyl-Heisenberg group.

\section{The Weyl-Heisenberg and Discrete Symplectic Groups}
We begin by enumerating some salient facts concerning the Weyl-Heisenberg and Discrete Symplectic groups which will be needed in the sequel.  For the proofs of the statements which follow the reader may consult, for example, refs.~\cite{me1,ADF}.

To define the Weyl-Heisenberg group choose an orthonormal basis $|0\rangle, \dots |d-1\rangle$ for $\mathcal{H}_d$ and define operators $X$ and $Z$ by
\begin{align}
X |a\rangle & = | a+1 \rangle
\label{eq:Xdef}
\\
Z |a\rangle & = \omega^{a} |a\rangle
\label{eq:Zdef}
\end{align}
for all $a$, where addition of indices is \emph{modulo} $d$ and where $\omega= e^{2 \pi i/d}$.  The Weyl-Heisenberg displacement operators are then defined by
\be
D_{\mathbf{q}} = \omega^{2^{-1} q_1 q_2} X^{q_1} Z^{q_2}
\ee
where the integers $q_1, q_2$ range over the interval $[0,d-1]$ and
where $2^{-1}$ denotes the unique \emph{integer} in the range $[0,d-1]$ such that $2^{-1} \times 2 = 1 \text{ mod $d$}$ (so, for example, $2^{-1} = 3$ if $d=5$).  Note that these definitions depend on the fact that $d$ is odd; for even $d$ the definition of the displacement operators is a little more complicated.

We  will also have occasion to use the discrete symplectic group $\SL(2,\mathbb{Z}_d)$, consisting of all  matrices of the form
\be
F = \bmt \alpha & \beta \\ \gamma & \delta \emt
\ee
where $\alpha, \beta, \gamma, \delta$ are integers \emph{modulo} $d$ such that $\det F = 1 \text{ mod $d$}$ (and where we are again making use of some simplifications due to the fact that $d$ is odd).  For given $F$ define a corresponding unitary operator $U_F$ by~\cite{me1}
\be
U_F 
=
\begin{cases}
\sqrt{\frac{1}{d}} \sum_{a,b=0}^{d-1} \omega^{(2 \beta)^{-1} \left(\alpha b^2 - 2 a b + \beta b^2\right)} 
|a \rangle \langle b | \qquad & \beta \neq 0 
\\
\sum_{a=0}^{d-1} \omega^{2^{-1} \alpha \gamma a^2} |\alpha a \rangle \langle a | 
\qquad & \beta = 0
\end{cases}
\ee
(where, as before, the notation $(2\beta)^{-1}$, $2^{-1}$ signifies the inverse in the \emph{modular} sense; it should also be noted that we are making use of the fact that $d$ is specifically an odd \emph{prime}---if $d$ is non-prime the definition of $U_F$ is a little more complicated).
Then
\be
U^{\vphantom{\dagger}}_F D_{\mathbf{q}} U^{\dagger}_F = D^{\vphantom{\dagger}}_{F\mathbf{q}}
\ee
for all $\mathbf{p}$.  

We can obtain a full set of MUBs by acting on the standard basis $|0\rangle, \cdots , | d-1\rangle$ with symplectic unitaries $U_F$.  For the sake of definiteness define
\begin{align}
G & = \bmt 1 & 1 \\ 0 & 1 \emt \\
H & = \bmt 0 & 1 \\ -1 & 0 \emt
\end{align}
and
\be
|m, a\rangle = \begin{cases}
U_G^m |a\rangle \qquad & m = 0, 1, \dots, d-1 \\
U_H |a\rangle \qquad & m = \infty
\end{cases}
\ee
Then it is readily verified that the bases $| m, 0\rangle, |m, 1\rangle , \dots | m, d-1\rangle$ as $m$ ranges over the set $\{0, 1, \dots (d-1),\infty\}$ constitute a full set of $d+1$ MUBs.  Note that the basis for which $m=0$ is just the standard basis.  The choice of the label $m=\infty$ rather than $m=d$ to denote the last basis is motivated by the discrete affine plane picture~\cite{WG}.  It is shown in ref.~\cite{ADF} that this is the only set of MUBs which can be obtained by acting on the standard basis with symplectic unitaries $U_F$ (up to permutation and re-phasing).  To describe the action of the Weyl-Heisenberg displacement operators it is convenient first to define operators $X_m$, $Z_m$ by
\begin{align}
X_m | m, a\rangle &= | m, a+1\rangle
\\
\intertext{and}
Z_m | m, a\rangle & = \omega^{2^{-1} a} |m, a\rangle
\end{align}
(\emph{c.f} Eqs.~(\ref{eq:Xdef}) and~(\ref{eq:Zdef})).  In terms of these operators we then have
\be
D_{\mathbf{q}} | m, a\rangle
=
\begin{cases}
\omega^{2^{-1} (q_1-m q_2)q_2} X_m^{q_1-mq_2} Z_m^{q_2} | m,a \rangle  \qquad & m \neq \infty
\\
\omega^{-2^{-1} q_1 q_2} X_m^{-q_2} Z^{q_1}_m | m, a \rangle \qquad & m =\infty
\end{cases}
\label{eq:DisOpOnMUBs}
\ee
(in other words the Weyl-Heisenberg displacement operators act on each basis by permuting and rephasing).

A SIC-POVM covariant under the action of the Weyl-Heisenberg group is defined by first choosing a fiducial vector $|\psi\rangle$ with the property
\be
\left| \langle \psi | D_{\mathbf{q}} | \psi \rangle \right|^2
=\frac{1}{d+1} \left( d \delta_{\mathbf{q},\boldsymbol{0}} + 1\right)
\ee
for all $\mathbf{q}$ (assuming this to be possible), and then defining
\be
|\psi_{\mathbf{q}} \rangle = D_{\mathbf{q}} | \psi\rangle
\ee
\section{Geometrical Relationships of the SIC and MUB Simplices}
These preliminaries concluded we are now in a position to analyze the geometrical relationships between a single SIC simplex and the $d+1$ MUB simplices (in odd prime dimension, for Weyl-Heisenberg covariant SICs and MUBs).

Let $\mathbf{B}^{\mathrm{m}}_{m,a}$ be the Bloch vector corresponding to  $|m,a\rangle\langle m,a |$, and let $\mathbf{B}^{\mathrm{s}}_{\mathbf{q}}$ be the Bloch vector corresponding to $|\psi_{\mathbf{q}}\rangle\langle \psi_{\mathbf{q}}|$.  So~\cite{Ingemar1,Ingemar2,me2}
\begin{align}
\langle \mathbf{B}^{\mathrm{m}}_{m,a}, \mathbf{B}^{\mathrm{m}}_{m',a'}\rangle
& =
\begin{cases}
\frac{1}{d-1} \left(d \delta_{aa'} -1\right) \qquad & m = m' \\
0 \qquad & m \neq m'
\end{cases}
\\
\intertext{and}
\langle \mathbf{B}^{\mathrm{s}}_{\mathbf{q}}, \mathbf{B}^{\mathrm{s}}_{\mathbf{q}'}\rangle
& = 
\frac{1}{d^2-1} \left( d^2 \delta_{\mathbf{q},\mathbf{q}'} - 1\right)
\end{align}
Let $\mathcal{P}_{m}$ denote the subspace spanned by the vectors $\mathbf{B}^{\mathrm{m}}_{m,0}, \dots \mathbf{B}^{\mathrm{m}}_{m,d-1}$.
It will also be convenient to choose an orthonormal basis for Bloch space.  Let $B^{\mathrm{m}}_{m,a,u}$ and $B^{\mathrm{s}}_{\mathbf{q},u}$ be the components of $\mathbf{B}^{\mathrm{m}}_{m,a}$ and $\mathbf{B}^{\mathrm{s}}_{\mathbf{q}}$ respectively relative to this basis (where the index $u$ runs from $1$ to $d^2-1$).

Now define $Q_m$ to be the matrix with elements
\be
Q_{m,uu'} = \frac{d-1}{d} \sum_{a = 0}^{d-1} B^{\mathrm{m}}_{m,a, u} B^{\mathrm{m}}_{m,a, u'}
\ee
It is easily seen that $Q_m$ is the projection operator projecting onto the $m^{\mathrm{th}}$ MUB hyperplane  $\mathcal{P}_m$:
\be
Q_m \mathbf{B}^{\mathrm{m}}_{m' ,a} =
\begin{cases}
\mathbf{B}^{\mathrm{m}}_{m,a} \qquad & m'=m
\\
0 \qquad & m' \neq m
\end{cases}
\ee
We wish to calculate the projections of the Bloch vectors $\mathbf{B}^{\mathrm{s}}_{\mathbf{q}}$.  We have
\begin{align}
Q_m \mathbf{B}^{\mathrm{s}}_{\mathbf{q}}
&=\frac{d-1}{d} \sum_{a=0}^{d-1} \langle \mathbf{B}^{\mathrm{m}}_{m,a},\mathbf{B}^{\mathrm{s}}_{\mathbf{q}} \rangle \mathbf{B}^{\mathrm{m}}_{m,a}
\nonumber
\\
& = \frac{1}{d^2} \sum_{a=0}^{d-1} \tr\left( \mathbf{B}^{\mathrm{m}}_{m,a}\mathbf{B}^{\mathrm{s}}_{\mathbf{q}} \right)\mathbf{B}^{\mathrm{m}}_{m,a}
\end{align}
Now
\begin{align}
 \tr\left( \mathbf{B}^{\mathrm{m}}_{m,a}\mathbf{B}^{\mathrm{s}}_{\mathbf{q}} \right)
 & = 
 \tr \Bigl(\bigl( d |m,a\rangle\langle m,a | -1\bigr) \bigl(d |\psi_{\mathbf{q}}\rangle\langle \psi_{\mathbf{q}} | -1\bigr)
 \Bigr)
 \nonumber
 \\
 & =
 d^2 \bigl| \langle m, a | \psi_{\mathbf{q}}\rangle \bigr|^2 - d
 \nonumber
 \\
 & = 
 d^2 \bigl| \langle m, a | D_{\mathbf{q}} | \psi \bigr|^2 - d
 \nonumber
 \\
 & =
 \begin{cases}
 d^2 p_{m,(a-q_1+ m q_2)} -d \qquad & m \neq \infty
 \\
 d^2 p_{m,(a+ q_2)} -d \qquad & m = \infty
 \end{cases}
\end{align}
where we used Eq.~(\ref{eq:DisOpOnMUBs}) in the last line, and where
\be
p_{m,a} = \bigl| \langle m,a| \psi\rangle \bigr|^2
\ee
is the probability of obtaining outcome $a$ when measuring the $m^{\mathrm{th}}$ MUB in the fiducial state $|\psi\rangle$.  Taking into account the fact that
\be
\sum_{a=0}^{d-1} \mathbf{B}^{\mathrm{m}}_{m,a} = 0
\ee
for all $m$ we deduce
\be
Q_m \mathbf{B}^{\mathrm{s}}_{\mathbf{q}}
=
\begin{cases}
\frac{1}{\sqrt{d+1}} \mathbf{C}_{m,(q_1-m q_2)} \qquad & m \neq \infty
\\
\frac{1}{\sqrt{d+1}} \mathbf{C}_{m,-q_2} \qquad & m = \infty
\end{cases}
\ee
where
\be
\mathbf{C}_{m,a} = \sqrt{d+1} \sum_{b=0}^{d-1} p_{m,b} \mathbf{B}_{m,b+a}
\ee
It was shown in ref.~\cite{ADF} that 
\be
\sum_{b=0}^{d-1} p_{m,b} p_{m,b+a} = \frac{1}{d+1}\left(\delta_{a,0}+1\right)
\ee
Consequently
\be
\langle \mathbf{C}_{m,a} , \mathbf{C}_{m,a'} \rangle 
 = \frac{1}{d-1} \left( d \delta_{a,a'} -1\right)
\ee
from which it can be seen that the vectors $\mathbf{C}_{m,a}$ constitute a $d-1$ dimensional regular simplex in the MUB hyperplane $\mathcal{P}_m$.   This gives us two regular simplices in each MUB hyperplane:  the one comprising the vectors $\mathbf{B}^{\mathrm{m}}_{m,a}$, and the one comprising the vectors $\mathbf{C}^{\vphantom{\mathrm{m}}}_{m,a}$.

Putting all this together we conclude
\begin{enumerate}
\item The vector $\mathbf{B}^{\mathrm{s}}_{\mathbf{q}}$ makes the same angle with each of the hyperplanes $\mathcal{P}_m$.   To put it another way:  the Bloch vectors representing the SIC-POVM are maximally distant from the MUB hyperplanes.  
\item  For each $m$, the $d^2$ vectors $\mathbf{B}^{\mathrm{s}}_{\mathbf{q}}$ constituting the SIC-simplex project onto only $d$ vectors.   More specifically, if $m \neq \infty$ the $d$ vectors $\{\mathbf{B}^{\mathrm{s}}_{(a+mb,b)}\colon  b = 0, 1, \dots , d-1\}$ project onto the single vector $(d+1)^{-1/2}\mathbf{C}_{m,a}$, while if $m=\infty$ the $d$ vectors $\{\mathbf{B}^{\mathrm{s}}_{(b,-a)}\colon  b = 0, 1, \dots , d-1\}$ project onto the single vector $(d+1)^{-1/2}\mathbf{C}_{m,a}$.
\item For each $m$ the $d-1$ dimensional projection of the $d^2-1$ dimensional SIC simplex  is itself a regular simplex.
\end{enumerate}

Since the vectors $\mathbf{B}^{\mathrm{m}}_{m,a}$ and $\mathbf{C}_{m,a}$ both constitute regular simplices of the same size it must be possible to rotate one onto the other.  We now investigate this rotation.

Let $R_m$ be the matrix with elements
\be
R_{m,uv} = \frac{d-1}{d} \sum_{a=0}^{d-1} C_{m,a,u} B^{\mathrm{m}}_{m,a,v}
\ee
It is easily seen that  
\begin{align}
R^{\mathrm{T}}_{m} R^{\vphantom{\mathrm{T}}}_{m}
 &= Q_m
 \\
 \intertext{and}
 R_m \mathbf{B}^{\mathrm{m}}_{m,a}& = \mathbf{C}^{\vphantom{\mathrm{m}}}_{m,a}
\end{align}
So the restriction of $R_m$ to $\mathcal{P}_m$ is a real orthogonal transformation taking the $\mathbf{B}^{\mathrm{m}}_{m,a}$ simplex onto the $\mathbf{C}^{\vphantom{\mathrm{m}}}_{m,a}$ simplex.

We next diagonalize $R_m$, regarded as a complex matrix.  Define
\be
\tilde{\mathbf{B}}^{\mathrm{m}}_{m,a} = \frac{\sqrt{d-1}}{d} \sum_{b=0}^{d-1} \omega^{a b} \mathbf{B}^{\mathrm{m}}_{m,b}
\ee
for $a=1, \dots , d-1$ (the definition does, of course, also make sense for $a=0$; however $\tilde{\mathbf{B}}^{\mathrm{m}}_{m,0} = 0$).  It is straightforward to verify
\begin{align}
\mathbf{B}^{\mathrm{m}}_{m,a}
&= \frac{1}{\sqrt{d-1}} \sum_{b=1}^{d-1} \omega^{-ab} \tilde{\mathbf{B}}^{\mathrm{m}}_{m,b}
\\
\intertext{(where $a$ runs from $0$ to $d-1$) and}
\sum_{u=0}^{d-1} \left( \tilde{B}^{\mathrm{m}}_{m,a,u}\right)^{*}\tilde{B}^{\mathrm{m}}_{m,b,u}
& = \delta_{a,b}
\end{align}
(where $a,b$ run from $1$ to $d-1$).
So the vectors $\tilde{\mathbf{B}}^{\mathrm{m}}_{m,a}$ are an orthonormal basis for $\mathcal{P}_m$, regarded as a complex vector space. They are also eigenvectors of $R_m$:
\begin{align}
\left(R^{\vphantom{m}}_m \tilde{\mathbf{B}}^{\mathrm{m}}_{m,a}\right)_u
& =
\frac{(d-1)^{\frac{3}{2}}}{d^2} \sum_{b,c,v=0}^{d-1}\omega^{ac} C^{\vphantom{m}}_{m,b,u} B^{\mathrm{m}}_{m,b,v}
B^{\mathrm{m}}_{m,c,v}
\nonumber
\\
& =
\frac{\sqrt{d-1}}{d} \sum_{b=0}^{d-1} \omega^{ab} C_{m,b,u}
\nonumber
\\
& = 
\frac{\sqrt{d^2-1}}{d} \sum_{b,c=0}^{d-1} \omega^{ab} p_{m,c} B_{m,c+b,u}
\nonumber
\\
& =
\frac{\sqrt{d+1}}{d}\sum_{b,c=0}^{d-1} \sum_{e=1}^{d-1}\omega^{b(a-e)-ec} p^{\vphantom{m}}_{m,c} \tilde{B}^{\mathrm{m}}_{m,e,u}
\nonumber
\\
& = \tilde{p}_{m,a} \tilde{B}^{\mathrm{m}}_{m,a,u}
\nonumber
\\
\intertext{where}
\tilde{p}_{m,a} &= \sqrt{d+1} \sum_{b=0}^{d-1} \omega^{-a b} p_{m,b}
\end{align}
Regarded as a complex matrix $R_m$ is unitary, so we can write
\be
\tilde{p}_{m,a} = e^{i \theta_{m,a}}
\ee
for suitable phase angles $\theta_{m,a}$.  The fact that $\tilde{p}_{m,a}^{*} = \tilde{p}^{\vphantom{*}}_{m,-a}$ means we can assume 
\be
\theta_{m,-a} = - \theta_{m,a}
\ee
  for all $m$, $a$.  The matrix elements of $R_m$ can then be written
\be
R_{m,uv} = \sum_{a=1}^{\frac{d-1}{2}}
 \left(e^{i \theta_{m,a}} \tilde{B}^{\mathrm{m}}_{m,a,u} \left(\tilde{B}^{\mathrm{m}}_{m,a,v} \right)^{*}
 + e^{-i \theta_{m,a}} \tilde{B}^{\mathrm{m}}_{m,-a,u} \left(\tilde{B}^{\mathrm{m}}_{m,-a,v} 
\right)^{*}\right)
\ee
We can now use this expression to write $R_m$ explicitly as a rotation matrix in canonical form.  Define 
\begin{align}
\tilde{\mathbf{B}}^{\mathrm{R}}_{m,a} &= \frac{1}{\sqrt{2}} \bigl(\tilde{\mathbf{B}}_{m,-a} + \tilde{\mathbf{B}}_{m,a}\bigr)
=\frac{\sqrt{2(d-1)}}{d} \sum_{b=0}^{d-1} \cos\left(\frac{2 a b \pi i}{d}\right) \mathbf{B}^{\mathrm{m}}_{m,b}
\\
\tilde{\mathbf{B}}^{\mathrm{I}}_{m,a} &= \frac{i}{\sqrt{2}} \bigl(\tilde{\mathbf{B}}_{m,-a} -\tilde{\mathbf{B}}_{m,a}\bigr)
=\frac{\sqrt{2(d-1)}}{d} \sum_{b=0}^{d-1} \sin\left(\frac{2 a b \pi i}{d}\right) \mathbf{B}^{\mathrm{m}}_{m,b}
\end{align}
for $a=1, \dots, (d-1)/2$.  By  construction the vectors $\tilde{\mathbf{B}}^{\mathrm{R}}_{m,a}, \tilde{\mathbf{B}}^{\mathrm{I}}_{m,a}$ are a \emph{real} orthonormal basis for Bloch space:
\begin{align}
\langle \tilde{\mathbf{B}}^{\mathrm{R}}_{m,a},\tilde{\mathbf{B}}^{\mathrm{R}}_{m,b}\rangle
& = \langle \tilde{\mathbf{B}}^{\mathrm{I}}_{m,a},\tilde{\mathbf{B}}^{\mathrm{I}}_{m,b}\rangle
=\delta_{a,b}
\\
\langle \tilde{\mathbf{B}}^{\mathrm{R}}_{m,a},\tilde{\mathbf{B}}^{\mathrm{I}}_{m,b}\rangle
& = 0
\end{align}
for $a,b=1, \dots, (d-1)/2$.
Moreover
\begin{align}
R_{m,uv} & =
\sum_{a=1}^{\frac{d-1}{2}}\biggl( \cos \theta_{m,a} \Bigl(\tilde{B}^{\mathrm{R}}_{m,a,u}\tilde{B}^{\mathrm{R}}_{m,a,v}
+ \tilde{B}^{\mathrm{I}}_{m,a,u}\tilde{B}^{\mathrm{I}}_{m,a,v}\Bigr)
\nonumber
\\
& \hspace{1 in}
+\sin \theta_{m,a} \Bigl(\tilde{B}^{\mathrm{I}}_{m,a,u}\tilde{B}^{\mathrm{R}}_{m,a,v}
- \tilde{B}^{\mathrm{R}}_{m,a,u}\tilde{B}^{\mathrm{I}}_{m,a,v}\Bigr) \biggr)
\end{align}
So when it is expressed in terms of the basis $\tilde{\mathbf{B}}^{\mathrm{R}}_{m,1}, \tilde{\mathbf{B}}^{\mathrm{I}}_{m,1},  \dots, \tilde{\mathbf{B}}^{\mathrm{R}}_{m,(d-1)/2}, \tilde{\mathbf{B}}^{\mathrm{I}}_{m,(d-1)/2}$ the operator $R_m$ is the direct sum of the $2\times 2$ rotation matrices
\be
\bmt \cos\theta_{m,a} & - \sin \theta_{m,a} \\ \sin \theta_{m,a} & \cos \theta_{m,a} \emt
\ee

It follows from all this that if we knew the angles $\theta_{m,a}$ then we could easily reconstruct the SIC-POVM.  All we would need to do is to use the operators $R_m$ to rotate the MUB simplices $\mathbf{B}^{\mathrm{m}}_{m,a}$ onto the projected SIC simplices $\mathbf{C}^{\vphantom{m}}_{m,a}$.  We could then calculate the SIC-POVM using the formula
\be
\mathbf{B}^{\mathrm{s}}_{\mathbf{q}}
=
\frac{1}{\sqrt{d+1}} \left(\mathbf{C}_{\infty,-q_2}+ \sum_{m=0}^{d-1} \mathbf{C}_{m,(q_1-mq_2)} \right) 
\ee

When we embarked on this investigation we entertained the (faint) hope that the angles $\theta_{m,a}$ would turn out to have a simple form (for instance, that they would all be rational multiples of $\pi$).  Unfortunately this does not seem to be the case.  So the result does not seem to take us any closer to solving the existence problem.  Nevertheless, it does, perhaps, have some intrinsic interest.  
\section{Minimum Uncertainty States and the Fiduciality Condition}
The results in the last section provide some insight into the geometrical significance of the minimum uncertainty condition introduced by Sussman and Wootters~\cite{WS,Suss} and Appleby, Dang and Fuchs~\cite{ADF}.  It was shown by these authors that a pure state $|\psi\rangle$ minimizes the quadratic Renyi entropy averaged over a full set of MUBs if and only if
\be
\sum_{a = 0}^{d-1} p^{2}_{m,a} = \frac{2}{d+1}
\ee
for all $m$, where $p_{m,a} = |\langle m, a | \psi \rangle|^2$.  

Let $\mathbf{B}$ be the Bloch vector corresponding to an arbitrary normalized state $|\psi\rangle$.  Then 
\be
Q_m \mathbf{B} = \sum_{a=0}^{d-1} p_{m,a} \mathbf{B}^{\mathrm{m}}_{m,a}
\ee
Consequently
\be
\| Q_m \mathbf{B} \|^2 = \frac{d}{d-1} \left( \sum_{a=0}^{d-1} p^{2}_{m,a} \right) - \frac{1}{d-1}
\ee
from which it follows that 
\be
\| Q_m \mathbf{B} \|^2 = \frac{1}{d+1}
\ee
for all $m$ if and only if
\be
\sum_{a = 0}^{d-1} p^{2}_{m,a} = \frac{2}{d+1}
\ee
for all $m$.  In other words $|\psi\rangle$ is a minimum uncertainty state if and only if its Bloch vector makes the same angle with all the MUB hyperplanes.

Appleby, Dang and Fuchs~\cite{ADF} further show that  $|\psi\rangle$ is a fiducial state for a Weyl-Heisenberg covariant SIC-POVM if and only if it is a minimum uncertainty state and, in addition,
\be
\sum_{a=0}^{d-1} p_{m,a}p_{m,a+b} = \frac{1}{d+1}
\label{eq:ADFFidCondB}
\ee
for all $m$ and all $b \neq 0 \text{ mod $d$}$.  This statement also has a simple geometrical interpretation, as we now show.

For each $m$ define $S_m$ to be the operator whose matrix elements are
\be
S_{m,uv} = \frac{d-1}{d} \sum_{a=0}^{d-1} B^{\mathrm{m}}_{m,a+1,u} B^{\mathrm{m}}_{m,a,v}
\ee
It is readily confirmed that 
\be
S^{\mathrm{T}}_{m} S^{\vphantom{T}}_{m} = Q^{\vphantom{T}}_m
\ee
So the restriction of $S_m$ to the $m^{\mathrm{th}}$ MUB hyperplane $\mathcal{P}_m$ is an orthogonal operator.  It is in fact a rotation operator (as follows from the fact that $S^{d}_m = Q^{\vphantom{d}}_m$ which, since $d$ is odd, means $\det (S_m) = 1$). Its significance is that it effects the action of the Weyl-Heisenberg displacement operators on $\mathcal{P}_m$:
\be
D_{\mathbf{q}} Q_m \mathbf{B} D^{\dagger}_{\mathbf{q}}
=
\begin{cases}
S_m^{q_1-m q_2 } Q_m \mathbf{B} \qquad & m \neq \infty
\\
S_m^{-q_2} Q_m \mathbf{B} \qquad & m = \infty
\end{cases}
\ee
Now let $\mathbf{B}$ be the Bloch vector corresponding to a minimum uncertainty state.  We have
\be
Q_m \mathbf{B} = \frac{d-1}{d} \sum_{a=0}^{d-1} \langle \mathbf{B}^{\mathrm{m}}_{m,a},\mathbf{B}^{\vphantom{m}} \rangle \mathbf{B}^{\mathrm{m}}_{m,a}
=
\sum_{a=0}^{d-1} p^{\mathrm{m}}_{m,a} \mathbf{B}^{\mathrm{m}}_{m,a}
\ee
and, consequently,
\be
\langle Q^{\vphantom{a}}_m \mathbf{B}, S^{a}_{m} Q^{\vphantom{a}}_m \mathbf{B} \rangle
= \frac{d}{d-1} \sum_{b=0}^{d-1} p_{m,b} p_{m,b+a} - \frac{1}{d-1}
\ee
So eq.~(\ref{eq:ADFFidCondB}) is satisfied if and only if
\be
\langle Q^{\vphantom{a}}_m \mathbf{B}, S^{a}_{m} Q^{\vphantom{a}}_m \mathbf{B} \rangle
=
\frac{1}{d^2-1} \left( d \delta_{a,0} - 1\right)
\ee
(\emph{i.e.}\ if and only if the vectors $S^{a}_{m} Q^{\vphantom{a}}_m \mathbf{B}$ form a $d-1$ dimensional regular simplex for each fixed $m$).

To summarise:  a state $|\psi\rangle$ with Bloch vector $\mathbf{B}$ is a fiducial state for a Weyl-Heisenberg SIC-POVM if and only if
\begin{enumerate}
\item $\mathbf{B}$ makes the same angle with each MUB hyperplane
\item For each $m$ the rotated vectors $Q^{\vphantom{a}}_m \mathbf{B}, S^{\vphantom{a}}_m Q^{\vphantom{a}}_m \mathbf{B}, \dots, S^{d-1}_m Q^{\vphantom{a}}_m \mathbf{B}$ form a $d-1$ dimensional regular simplex.
\end{enumerate}
\subsection*{Acknowledgments}
This research was supported in part by the U. S. Office of Naval Research (Grant No. N00014-09-1-0247) and by the Perimeter Institute for Theoretical Physics.

\end{document}